\documentclass[10pt,aps,amsmath,showpacs,amssymb,pre,twocolumn]{revtex4}
\usepackage[dvips]{graphicx}
\usepackage[dvips,usenames]{color}
\usepackage{amsmath}

\newcommand{\av}[1]{\langle {#1} \rangle}

\begin{document}

\title{Rare regions of the Susceptible Infected Susceptible model on Barab\'asi-Albert networks} 

\author{G\'eza \'Odor}
\affiliation{Research Centre for Natural Sciences, 
Hungarian Academy of Sciences, MTA TTK MFA, 
P. O. Box 49, H-1525 Budapest, Hungary}

\date{\today}

\begin{abstract}
I extend a previous work to Susceptible-Infected-Susceptible (SIS) models on 
weighted Barab\'asi-Albert scale-free networks. Numerical evidence is provided 
that phases with slow, power-law dynamics emerge as the consequence of quenched 
disorder and tree topologies studied previously with the Contact Process. 
I compare simulation results with spectral analysis of the networks and show 
that the quenched mean-field (QMF) approximation provides a reliable, relatively 
fast method to explore activity clustering. This suggests that QMF can be used 
for describing rare-region effects due to network inhomogeneities. 
Finite size study of the QMF shows the expected disappearance of the epidemic 
threshold $\lambda_c$ in the thermodynamic limit and an inverse participation ratio
$\sim 0.25$, meaning localization in case of disassortative weight scheme. 
Contrary, for the multiplicative weights and the unweighted trees this value 
vanishes in the thermodynamic limit, suggesting only weak rare region effects 
in agreement with the dynamical simulations.
Strong corrections to the mean-field behavior in case of disassortative weights 
explains the concave shape of the order parameter $\rho(\lambda)$ at the 
transition point. Application of this method to other models may reveal 
interesting rare-region effects, Griffiths Phases as the consequence of 
quenched topological heterogeneities. 
\end{abstract}

\maketitle

%%%%%%%%%%%%%%%%%%%%%%%%%%%%%%%%%%%%%%%%%%%%%%%%%%%%%%%%%%%%%%%%%%%%%%%%
\section{Introduction}
%%%%%%%%%%%%%%%%%%%%%%%%%%%%%%%%%%%%%%%%%%%%%%%%%%%%%%%%%%%%%%%%%%%%%%%%

The research of nonequilibrium models has been a central topic 
of statistical mechanics \cite{marro1999npt,odorbook,Henkel}. 
A fundamental dynamical model to understand them is the Contact Process (CP) 
\cite{harris74,liggett1985ips}, in which sites can be either active (infected) 
or inactive (susceptible).
By decreasing the infection rate of the neighbors $\lambda/k$, where $k$ is the
degree of the vertex, a continuous phase transition occurs at some $\lambda_c$ 
critical point from the active to the inactive steady state. 
The latter is also called absorbing, because no spontaneous activation of sites 
is allowed and the density of infection ($\rho$) is zero.

Recently the interest has been shifted from models, defined on Euclidean, 
regular lattices to processes living on general networks 
\cite{barabasi02,mendesbook}.
The effects of the heterogeneous topological structures
\cite{dorogovtsev07:_critic_phenom,barratbook} is not yet fully understood.
In particular in the case of ubiquitous scale-free (SF) networks 
\cite{Barabasi:1999}, 
exhibiting $P(k)\sim k^{-\gamma}$ degree distribution of the nodes
\cite{barabasi02,mendesbook} the location of the phase transition and the
singular behavior is still a debated issue. Numerical simulations
\cite{Castellano:2006,Castellano:2008,PhysRevE.83.066113,FFCR11} and
theoretical approaches based on the heterogeneous mean-field (HMF)
theory \cite{Castellano:2006,Castellano:2008,boguna09:_langev} show
strong effects of the network heterogeneity on the behavior of the CP 
defined on complex networks.

Although SF-s exhibit infinite topological dimension ($d$), defined by
$N\propto r^d$, where $N$ is the number of nodes within the (chemical) 
distance $r$, simple mean-field approximations cannot capture several 
important features.  Studies of the CP, as well as other processes 
\cite{dorogovtsev07:_critic_phenom,barratbook} have shown that quenched 
disorder in networks is relevant in the dynamical 
systems defined on top of them. Very recently it has been shown, 
\cite{GPCP,odor:172,Juhasz:2011fk} that generic slow (power-law, or 
logarithmic) dynamics is observable by simulating CP on networks with 
finite $d$.  This observation is relevant for recent developments in 
dynamical processes on complex networks such as the simple model of 
``working memory'' \cite{Johnson}, brain dynamics \cite{Chialvo}, 
social networks with heterogeneous communities  
\cite{Castello}, or slow relaxation in glassy systems \cite{Amir}.

Slow dynamics can be the consequence of bursty behavior of agents 
connected by small world networks resulting in memory effects \cite{KK11}.
An independent cause is related to arbitrarily large ($l<N$), active
rare-regions (RR) with long lifetimes: $\tau\propto\exp(l)$ 
in the inactive phase ($\lambda \le \lambda_c$). In the region 
$\lambda_c^0 < \lambda < \lambda_c$, where $\lambda_c^0$ is the
critical point of the impure system, a so called Griffiths Phase (GP) 
\cite{Griffiths,Vojta} develops, with algebraic density decay 
$\rho \propto t^{-\alpha}$, $\alpha$ being a non-universal exponent. 
At and below the $\lambda_c^0$ rare clusters may still form, with an 
over-average mean degree, which are locally supercritical at a given 
$\lambda$ and induce a slower-than-exponential but faster than power-law 
decay of the global density. 
In case of these ``weak rare-region effects'' numerical simulations 
\cite{Juhasz:2011fk} showed stretched-exponential decay, 
$\rho(t)\sim e^{-const\cdot t^y}$ with exponents $y$ varying from values 
close to $1$, for small $\lambda$, to very small values 
(converging to $0$) for $\lambda$ approaching the GP.
The rare-region induced dynamical behavior can be described  
by non-perturbative methods
\cite{PhysRevLett.59.586,sethna88,PhysRevB.54.3328,PhysRevLett.69.534}.

More recently the possibility of slow dynamics has been investigated
on Barab\'asi-Albert (BA) networks with $\gamma=3$, possessing infinite 
$d$ \cite{BAGP}.
Very extensive simulations showed linear leading order density decay 
and $\rho(\lambda,t\to\infty)\propto |\lambda-\lambda_c|$ behavior
with logarithmic corrections, in agreement with the HMF approximations.
It was also pointed out that in case of BA networks with one link per 
incoming node the epidemic propagation slows down, and nontrivial critical
density decay emerges with $\rho(t,\lambda_c) \propto t^{-\alpha}$,
$\alpha \simeq 0.5$. Furthermore, when $k$ dependent weighting was also
applied, suppressing hubs or making the network disassortative
GP-like regions could be observed in the simulations. A systematic
finite scaling study revealed that these power-laws saturate in the
$N\to\infty$ thermodynamic limit, suggesting smeared phase transitions 
\cite{Vojta}.
This can be understood via the possibility of embedding even infinite 
dimensional, active RR-s in networks with $d = \infty$, leading to 
finite steady state density: $\rho(t\to\infty, N\to\infty)$.
A numerical percolation analysis \cite{BAGP} has strengthened this 
view indeed.

I continue the work \cite{BAGP} to explore whether quenched mean-field 
(QMF) theory \cite{GDOM12} with spectral decomposition (SD) can help 
understanding rare-region effects in complex networks. The QMF differs 
from HMF by taking into account variations of the quenched network topology. 
I study the Susceptible Infected Susceptible (SIS) model \cite{SIS}, 
instead of the CP, because the infection rate is independent of $k$, 
thus the rate equation involves symmetric matrices and a spectral   
analysis can be done on an orthonormal basis with real, non-negative 
eigenvalues. The SIS model is a two state system, in which infected 
sites propagate the epidemic (venerate all neighbors) with rate 
$\lambda$, or recover with rate $1$.

This extension is far from trivial, because the epidemic transmitting 
capability of the infected nodes is higher than in CP, i.e. it is not 
normalized by the number of outgoing edges. Therefore, the emergence 
of localized rare-regions (RR) is more difficult. I provide extensive 
simulation results, showing numerical evidence for GP like regions 
with generic slow dynamics, similarly as in case of the CP. 
The QMF and SD analysis set up for these cases is in good agreement 
with the dynamical simulations.

%%%%%%%%%%%%%%%%%%%%%%%%%%%%%%%%%%%%%%%%%%%%%%%%%%%%%%%%%%%%%%%%%%%%%%%%%
\section{Network models}
\label{sec:model-definition}
%%%%%%%%%%%%%%%%%%%%%%%%%%%%%%%%%%%%%%%%%%%%%%%%%%%%%%%%%%%%%%%%%%%%%%%%%

I consider here SIS on BA networks, in particular for loop-less 
and weighted cases as described in \cite{BAGP}.  
This permits very simple and fast construction, in contrast with
other standard network generation models, e.g \cite{ucmmodel}.
The BA growth starts with a fully connected graph of size $N_0 = 10$ nodes, 
but comparisons with $N_0 = 5$ and $20$ have also been done to
test any dependence.
For BA at each time step $s$, a new vertex (labeled by $s$) with $m$ 
edges is added to the network and connected to an existing 
vertex $s'$ of degree $k_{s'}$ with probability 
$\Pi_{s  \rightarrow s'} = k_{s'} /\sum_{s''= 1}^{s''<s} k_{s''}$. 
This process is iterated until reaching the desired network size $N$. 
The resulting network has a SF degree distribution $P(k) \simeq k^{-3}$.
For $m=1$ we obtain a BA tree (BAT) topology, while for the looped case
$m=3$ was applied.

Binary (non-weighted) BA networks have been transformed to weighted
ones by assigning to every edge connecting vertexes $i$ and $j$ a
symmetric weight $\omega_{ij}$. In \cite{BAGP} two different network 
topology dependent quenched weight assignment strategy was introduced
in order to slow down and localize epidemics.

(i) \emph{Weighted BA tree I (WBAT-I):} Multiplicative weights, 
suppressing the infection capability of highly connected nodes
\begin{equation}
  \label{eq:2}
  \omega_{ij} = \omega_0 (k_i k_j)^{-\nu},
\end{equation}
where $\omega_0$ is an arbitrary scale and $\nu$ is a characteristic
exponent with $\nu\geq0$. This can model internal limitations of hubs,
like the sub-linear Heap's law \cite{Heap}.
In this paper I study the case with $\nu=1.5$ exponent.

(ii) \emph{Weighted BA tree II (WBAT-II):} Disassortative weighting
scheme according to the age of nodes in the network construction
\begin{equation}
  \label{eq:3}
  \omega_{ij} =\frac{|i-j|^x}{N},
\end{equation}
where the node numbers $i$ and $j$ correspond to the time step
when they were connected the network. Since the degree of nodes 
decreases as $k_i\propto (N/i)^{1/2}$ during this process, 
this selection with $x>0$ favors connection between unlike 
nodes and suppresses interactions between similar ones.
In \cite{BAGP} simulations showed evidence for a phase with power-law
dynamics of the CP for $x=2,3$ with CP. 
This paper is concerned about $x=2$ networks.

The presence of these weights affects the dynamics of the SIS. 
Thus, the rate at which a healthy vertex $i$ becomes ill on contact 
with an infected (active) vertex $j$ is proportional to $\lambda \omega_{ij}$, 
therefore the epidemic can in principle become trapped in isolated 
connected subsets. These regions of size $l$ are rare in general: 
$P(l) \propto \exp(-l)$, but can exhibit exponentially long lifetimes 
$\tau(l) \propto \exp(l)$.
In the healthy (inactive) phase they provide the leading order contribution
to the density of infected sites
\begin{equation}
\rho(t) \sim \int l P(l) \exp(-t/\tau) dl  \ ,
\end{equation}
which in the saddle point approximation results in $\rho(t)\sim t^{-\alpha}$ 
decay \cite{GPCP}.

%%%%%%%%%%%%%%%%%%%%%%%%%%%%%%%%%%%%%%%%%%%%%%%%%%%%%%%%%%%%%%%%%%
\section{Spectral analysis}
%%%%%%%%%%%%%%%%%%%%%%%%%%%%%%%%%%%%%%%%%%%%%%%%%%%%%%%%%%%%%%%%%%

In \cite{BAGP} the heterogeneous mean-field (HMF) analysis of CP
was worked out for these network models. However, extensive 
simulations showed different dynamical behaviors, except from the 
looped BA CP model case. Since HMF can't take into account rare 
region effects, nor models on trees that conclusion was not very
surprising. 

A mean-field theory, capable of describing the effects of 
quenched topologies of the network on which SIS is defined
is expected to give better agreement with numerical simulations.
The quenched mean-field (QMF) approach is based on the rate equation 
for $\rho_i(t)$, the infection probability of node $i$ at time 
$t$ \cite{GDOM12}:  
\begin{equation}
\label{qmfsis}
\frac{d\rho_i(t)}{dt} = -\rho_i(t)+(1-\rho_i(t))\sum_j A_{ij} \lambda \rho_j(t)~,
\end{equation}
where $A_{ij}$ is an element of the adjacency matrix assigned with $1$, 
if there is an edge between nodes $i$ and $j$ or $0$ otherwise.
This equation can be generalized by replacing the adjacency matrix
with the weighted adjacency matrix $B_{ij} = A_{ij} \omega_{ij}$, 
having weighted elements: $B_{ij}\in [0,1]$. 

For $t\rightarrow \infty$ the system evolves into a steady state, with the 
probabilities expressed as 
\begin{equation}
\rho_{i}=\frac{\lambda\sum_{j}B_{ij}\rho_{j}}{1+\lambda\sum_{j}B_{ij}\rho_{j}} \ .
\label{SIS2}
\end{equation}
Stability analysis shows that $\rho_{i} > 0$ above a $\lambda_{c}$ 
epidemic threshold, 
with finite order parameter (prevalence) $\rho \equiv \av{\rho_{i}}$ .

In the SD approach one expands $\rho_{i}$ in the space of eigenvectors 
of the adjacency matrix $A_{ij}$ (or $B_{ij}$ for weighted case) as
\begin{equation}
\rho_i=\sum_{\Lambda} c(\Lambda) f_{i} (\Lambda).
\label{exp}
\end{equation}
This extension can be done for real and symmetric weights, when the 
eigenvectors $\mbox{\boldmath$f$}(\Lambda)$ span a complete orthonormal basis.
For non-negative, symmetric matrices extension of the Perron-Frobenius
theorem asserts that it exhibits real eigenvalues, furthermore the largest one  
$\Lambda_{max} \equiv \Lambda_{1} \geq \Lambda_{2} \geq \dots \Lambda_{N}$
is unique and that the corresponding eigenvector 
$\mbox{\boldmath$f$}(\Lambda_1)$ has non-negative components.
In this basis Eq.~(\ref{SIS2}) can be expressed by the coefficients 
$c(\Lambda)$ as
\begin{equation}
c(\Lambda)=\lambda \sum_{\Lambda'} \Lambda' c(\Lambda') \sum_{i=1}^N \frac{f_i (\Lambda) f_i (\Lambda')}{1+\lambda \sum_{\widetilde{\Lambda}} \widetilde{\Lambda} c(\widetilde{\Lambda})f_i (\widetilde{\Lambda})}.
\label{SIS4}
\end{equation}
This gives $\lambda_c=1/\Lambda_1$ for the epidemic threshold and in its
neighborhood the order parameter can be approximated by the eigenvectors 
of the largest eigenvalues
\begin{equation}
\rho(\lambda) \approx a_1 \Delta + a_2 \Delta^2 + ... \ ,
\end{equation}
where $\Delta = \lambda \Lambda_{1}{-}1 {\ll} 1$ with the coefficients 
\cite{Mie} 
\begin{equation}
a_j = \sum_{i=1}^N f_i(\Lambda_j)/[N \sum_{i=1}^N f_i^3 (\Lambda_j)].
\label{epsilon}
\end{equation}
A homogeneous solution implies that a finite fraction of vertexes are 
infected right above $\lambda_c$ and $a_1$ is the order of $O(1)$.
That would mean that the components of $\mbox{\boldmath$f$}(\Lambda_1)$
are localized. On the other hand quenched inhomogeneous topologies can
also imply inhomogeneous $\rho_i$ distributions and one can assume that
they result in rare region effects, as in \cite{GPCP,odor:172,Juhasz:2011fk,BAGP}. 

To describe localization in the components of $\mbox{\boldmath$f$}(\Lambda_1)$
\cite{GDOM12} suggested to use the inverse participation ratio
\begin{equation}
IPR(\Lambda)\equiv \sum_{i=1}^{N} f_{i}^{4}(\Lambda),
\label{IPR}
\end{equation}
which in the limit $N \rightarrow \infty$ is of the order of $O(1)$ when 
the eigenvector is localized. Contrary, when $IPR(\Lambda){}\to{} 0$ 
this state is delocalized. Eq.~(\ref{epsilon}) implies that for
localized principal eigenvector $a_1 \sim O(1/N)$, thus
$\rho\approx a_1 \Delta \sim O(1/N)$, the disease is 
localized on a finite number $N\rho$ of vertexes. On the other hand,
when {\boldmath$f$}$(\Lambda_1)$ is delocalized the disease infects
a finite fraction of vertexes for $\lambda > \lambda_c$.

Goltsev et al. \cite{GDOM12} analyzed artificial and real SF networks
and found that in case of localized cases the epidemic threshold 
was actually absent and a real epidemic affecting a finite fraction 
of vertexes occurred after a smooth crossover at higher values of $\lambda$.
This is in agreement with the smeared phase transition scenario
proposed to explain the numerical results for CP on weighed BA
trees in \cite{BAGP}. In those systems rare-regions effects seemed
to arise, causing power-law density decays, which ultimately crossed
over to saturation to finite $\rho$-s in the $N\to\infty$ limit.
Right above $\lambda_c$ similar concave shaped $\rho(\lambda)$ was
obtained as in the localized cases of \cite{GDOM12}. 
Here I investigate the situation for SIS instead of CP
models, to see if this relation holds on. 
This has been done using finite size scaling analysis of the
quantities: $\lambda=1/\Lambda_1$, IPR and $a_i$  for $i=1,2,3$. 

It has been proven \cite{Chung} that for random, unweighted SF 
networks, with power-law degree distribution $P(k)\propto k^{-\gamma}$ 
the maximal eigenvalue scales as $\Lambda_1\propto\sqrt{k_{max}}$ 
for $\gamma>2.5$. Furthermore, the maximal degree of the network satisfies 
$k_{max}=\min\left[{N^{1/2},N^{1/(\gamma-1)}}\right]$, due to the structural 
cutoff of a finite network with $\gamma\leq3$.
Therefore, $\Lambda_1$ should scale with the network size as
\begin{equation}\label{Lscal}
\Lambda_1(N) \propto N^{1/4}
\end{equation}
for $\gamma=3$ considered here.

Using the software package OCTAVE I generated the $B_{ij}$ matrices of
BA, BAT, WBAT-I and WBAT-II networks for several sizes up to $N=200.000$ 
and calculated the three largest eigenvalues and the corresponding
eigenvectors. From these I deduced $1/\Lambda_1$, IPR and $a_i$. 
The whole SD analysis was done using the sparse matrix functions of 
OCTAVE to handle $B_{ij}$-s of the networks within reasonable computing times. 
For the largest sizes 1-2 weeks of a CPU time was needed.
Least-squares fitting with the form
\begin{equation} \label{lamscal}
1/\Lambda_1 = \lambda_c + X (1/N)^{c}
\end{equation}
has been applied for the largest eigenvalues.
As Table.~\ref{tabla} shows a good agreement was obtained with the 
finite size scaling expectation 
$\lim_{N\to\infty}1/\Lambda_1 = \lambda_c = 0$ 
of the QMF method.
The $N\to\infty$ extrapolated critical threshold values converge 
to zero using the three parameter fitting form (\ref{lamscal}). The fitted
power $c$ agrees with exponent of (\ref{Lscal}) reasonably well, except from
the WBAT-I case, where QMF results for only $N\le 6000$ could be achieved.

%%%%%%%%%%%%%%%%%%%%%%%%%%%%%%%%%%%%%%%%%%%%%%%%%%%%%%%%%%%%%%%%%%%%%%%%%
\begin{figure}[ht]
\includegraphics[height=6cm]{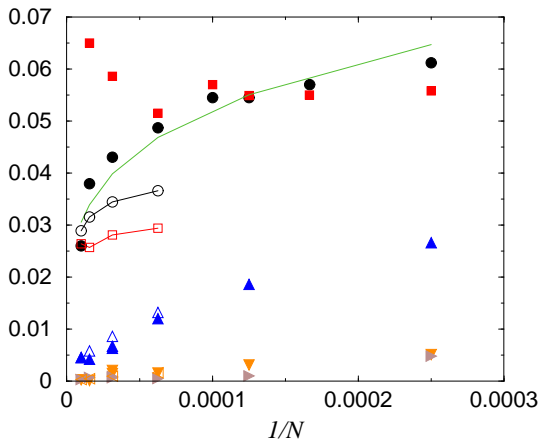}
\caption{\label{ba1} (Color online)
  Finite size scaling of QMF SD results for BAT model for
  $N=4000$, $8000$, $16000$, $32000$, $64000$, $100000$.
  Bullets: $\lambda_c$, boxes: IPR, up-triangles: $a_1$,
  down-triangles: $a_2$, right-triangles: $a_3$.
  Lines: least-squares fitting for with the form (\ref{lamscal}).
  Filled symbols correspond to $N_0=10$, hollow ones to $N_0=20$.
}
\end{figure}
%%%%%%%%%%%%%%%%%%%%%%%%%%%%%%%%%%%%%%%%%%%%%%%%%%%%%%%%%%%%%%%%%%%%%%%%

\begin{figure}[t]
\includegraphics[height=6cm]{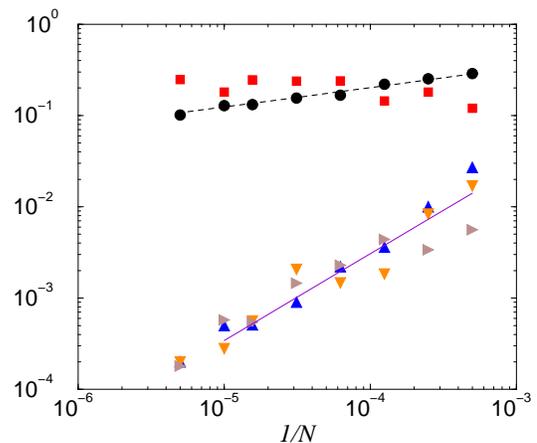}
\caption{\label{baw2} (Color online)
  Finite size scaling of QMF SD results for WBAT-II model for
  $N=2000$, $4000$, ... $200000$. Bullets: $\lambda_c$, boxes: IPR,
  up-triangles: $a_1$,  down-triangles: $a_2$, right-triangles: $a_3$.
  Lines: least-squares fitting for with the form (\ref{lamscal}).
}
\end{figure}

The IPR values decrease with $1/N$ and remain very small for the unweighted BA 
($IPR < 0.02$ ) and BAT ($IPR < 0.07$) networks, albeit in the latter case the 
tendency seems to change for $N \ge 32000$ as shown on Fig.~\ref{ba1}. 
This suggests an agreement with the anomalous scaling behavior of CP on 
the BA tree reported in \cite{BAGP}.
Here some clustering may be expected, which should be checked further 
by studying networks with $N >> 10^5$ nodes. 
Unfortunately this is out of the scope of the present study using OCTAVE, 
but is in preparation using high performance graphics cards.
For the WBAT-II case IPR stabilizes to $\simeq 0.23(2)$ 
(see Fig.~\ref{baw2}), advocating strong epidemic localization. 

One can speculate that the initial compact seeds, from which the BA graphs originate
can influence the clustering behavior of the epidemic in the steady state. 
By varying the seed size: $N_0 = 5, 10, 20$ no measurable effects were found 
on the WBAT results. For the unweighted BA and BAT networks on can see differences 
between the finite size results of the $N_0 =20$ and $N_0 = 10$ cases 
(see Fig.~\ref{ba1}), but the $N\to\infty$ asymptotic behavior appears to be the same. 

For the unweighted BA and BAT networks the eigenvalue analysis provides 
$a_1 >> a_2 >> a_3$ for all sizes in agreement with the $\beta=1$ HMF 
results for $\rho$ slightly above the critical threshold.
In the $N\to\infty$ limit $a_1(N) \sim (1/N)^{0.5}$ and 
$a_2 \simeq a_3 \simeq 0$ already for small sizes.
For the weighted WBAT-I case $a_1$ remains constant $\sim 0.35(1)$ while
$a_2 \sim a_3 \sim (1/N)$.
In case of the WBAT-II tree $a_i$ are of the same order of magnitude 
and vanish linearly with $1/N$ (see Fig.~\ref{baw2}), meaning a 
strong corrections to the leading order linear scaling. 
When one plots $\rho(\lambda)$ with these values one gets a concave 
curve from above and a tangential approach to $\lambda_c$.
Such steady state behavior has already been seen by simulations of CP on 
weighted BA networks \cite{BAGP}.
In the next section I compare these QMF results with simulations of SIS on
WBAT-I and WBAT-II networks.

%%%%%%%%%%%%%%%%%%%%%%%%%%%%%%%%%%%%%%%%%%%%%%%%%%%%%%%%%%%%%%%%%%%%%%%%%
%%%%%%%%%%%%%%%%%%%%%%%%%%%%%%%%%%%%%%%%%%%%%%%%%%%%%%%%%%%%%%%%%%%%%%%%

\begin{table}
\caption{Spectral QMF analysis results for SIS in different networks\label{tabla}}
\begin{center}
\begin{tabular}{|l|r|r|r|r|r|r|}
\hline
Network	& $\lambda$	& $c$ 	& 	IPR & $a_1$ 		& $a_2$		& $a_3$ \\
\hline
BA	& $0.02(3)$	& $0.32(4)$& $0.015(3)$ & $4\times10^{-4}$& $10^{-7}$	& $10^{-8}$   \\
BAT   	& $-0.001(2)$	& $0.24(1)$& $0.06(1)$ & $10^{-4}$	& $10^{-7}$ 	& $10^{-9}$ \\
WBAT-I  & $0.0001(3)$	& $0.9(1)$& $0.001(2)$ & $0.35(1)$	& $5\times10^{-8}$& $10^{-6}$ \\
WBAT-II & $0.001(1)$	& $0.23(2)$& $0.23(2)$ & $4\times 10^{-3}$& $3\times 10^{-5}$& $10^{-3}$ \\
\hline
\end{tabular}
\end{center}
\end{table}

%%%%%%%%%%%%%%%%%%%%%%%%%%%%%%%%%%%%%%%%%%%%%%%%%%%%%%%%%%%
\section{SIS model simulations on weighted trees}
\label{sec:sis-weight}
%%%%%%%%%%%%%%%%%%%%%%%%%%%%%%%%%%%%%%%%%%%%%%%%%%%%%%%%%%%

In simulations I considered the SIS model in continuous time as 
in \cite{FCP12}, to be in accordance with the rate equations.
At each time step a randomly chosen infected node recovers with  
probability $n_i / (n_i+\lambda n_n)$, where $n_i$ is the
the number of infected nodes and $n_n$ is the total number of links
emanating from them. Complementary, one of its randomly selected neighbor 
is infected, with probability $\lambda n_i / (n_i+\lambda n_n)$.
Following the reaction $N_i$ and $N_n$ are updated and the time is
incremented by $\Delta t = 1/ (n_i+\lambda n_n)$. The time is measured
by these Monte Carlo steps (MCs) and shown to be dimensionless on the figures.
These processes are iterated until $t < t_{max}$, or until the 
epidemic stops ($N_i=0$).

The networks were generated via the BA linear preferential attachment
rule \cite{Barabasi:1999}, following an initial fully connected seed of 
$N_0=20$. Neighbor indices of sites are stored in a dense matrix to 
save memory, thus up to $N=6\times 10^6$ sized networks could be studied.
The initial state was fully active and the concentration 
of infected sites was followed up to $t_{max} = 4\times 10^6$ MCs. 
Density decay runs were repeated and averaged over $\sim 10^4$ and
up to $10^5$ independent network realizations for WBAT-I and WBAT-II,
respectively. I have also calculated the effective decay exponents,
defined as the local slopes of $\rho(t)$ as given by
\begin{equation}  \label{aeff}
  \alpha_{\rm eff}(t) = - \frac {\ln[\rho(t)/\rho(t')]} 
  {\ln(t/t^{\prime})} \ ,
\end{equation}
where $t$ and $t^{\prime}$ have been chosen in such a way that the
discrete approximate of the derivative is sufficiently smooth. 
The static scaling behavior in the active steady state has also been 
investigated by measuring $\rho(\lambda,t\to\infty)$ deep in the 
saturation region.

%%%%%%%%%%%%%%%%%%%%%%%%%%%%%%%%%%%%%%%%%%%%%%%%%%%%%%%%%%%%%%%%%%%%%%%%%
\begin{figure}[t]
\includegraphics[height=6cm]{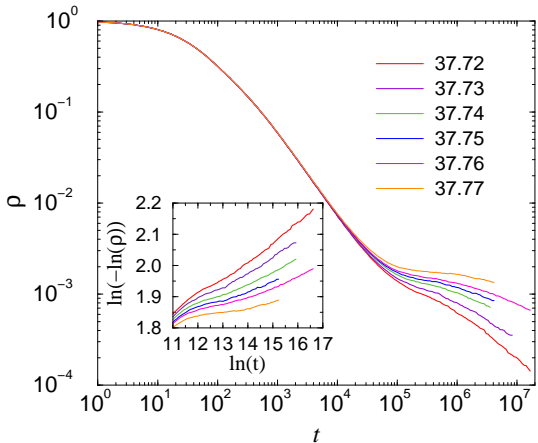}
\caption{\label{wbasisl} (Color online)
  Density decay as a function of time for the SIS model 
  on weighted BA trees generated with the WBAT-I scheme with exponent
  $\nu=1.5$. Network size $N=2\times 10^6$. Different curves correspond to
  $\lambda=$ 37.72, 37.73, 37.74, 37.75, 37.76, 37.77 (from bottom to top). 
  Inset: The same data plotted on the $\ln(-\ln(\rho(t)))$ vs. $\ln(t)$ 
  scale (with parameters from top to bottom order).}
\end{figure} 
%%%%%%%%%%%%%%%%%%%%%%%%%%%%%%%%%%%%%%%%%%%%%%%%%%%%%%%%%%%%%%%%%%%%%%%%

Figure~\ref{wbasisl} suggests that $\rho(t)$ of the WBAT-I model 
exhibits $\lambda$ dependent power-laws for $t > 3\times 10^5$ MCs.
This can be observed in networks with $N=2\times 10^6$ nodes
in the region $37.72 \le \lambda \le 37.77$.
The final slope of at the lowest, $\lambda = 37.72$ power-law curve is: 
$\alpha \simeq 0.343(4)$. By increasing the system size these curves 
become ultimately constant, as in case of the CP \cite{BAGP}, meaning that the
transition smears in the thermodynamic limit.
However, the above range is narrow both in $\lambda$ and the effective 
exponent $\alpha$. Furthermore, some down curvature can also be seen on the 
log.-log. plots in this region, which makes the power-law dynamics 
assumption questionable.
Since the QMF SD analysis suggests only weak rare-region effects I plotted 
the same data on the $\ln(-\ln(\rho(t)))$ versus $\ln(t)$ scale. 
As the inset of Fig.~\ref{wbasisl} shows straight lines appear asymptotically, 
corresponding to stretched exponential decay behavior.
In comparison for CP power-laws with $0.3 \le \alpha \le 1$ could be 
seen clearly in the region $140 \le \lambda \le 145$ \cite{BAGP}.

%%%%%%%%%%%%%%%%%%%%%%%%%%%%%%%%%%%%%%%%%%%%%%%%%%%%%%%%%%%%%%%%%%%%%%%%%
\begin{figure}[t]
\includegraphics[height=6cm]{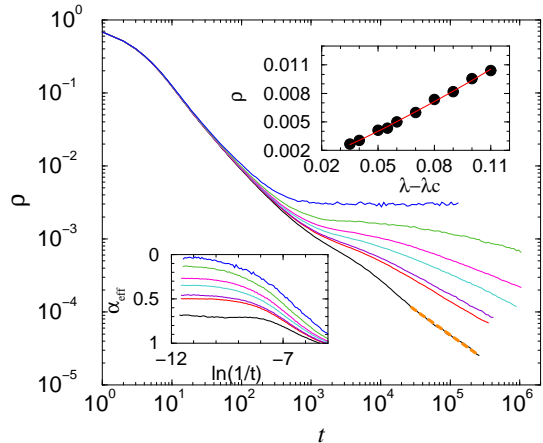}
\caption{\label{gwbasisl} (Color online)
  Density decay as a function of time for the SIS 
  on weighted BA trees generated with the age-dependent 
  (disassortative) WBAT-II scheme with exponent $x=2$. 
  Network size $N=10^6$. Different curves correspond to
  $\lambda=$ 1.657, 1.66, 1.661, 1.663, 1.67, 1.68, 1.69 
  (from bottom to top). Dashed line: power-law fit.
  Left inset: Local slopes of the same curves showing level-off 
  for large times. Right inset: Steady state density (bullets) 
  above the epidemic threshold. The line shows power-law fitting
  with the form (\ref{betafit}). 
}
\end{figure} 
%%%%%%%%%%%%%%%%%%%%%%%%%%%%%%%%%%%%%%%%%%%%%%%%%%%%%%%%%%%%%%%%%%%%%%%%

In the case of WBAT-II network power-laws decays of $\rho(t,\lambda)$ 
appear already for $t > 50.000$ MCs as shown in Fig.~\ref{gwbasisl}. 
To see corrections to scaling I have plotted the effective decay 
exponents (\ref{aeff}) in the left inset of Fig.~\ref{gwbasisl}.
The final slope of the lowest, $\lambda = 1.657$ curve tends to
$\alpha \simeq 0.70(2)$, which is far away from the HMF critical 
exponent value: $\alpha=1$. 
The dynamical scaling behavior appears in the $1.657 \le \lambda < 1.68$ 
region, at much lower infection rates than in case of the CP \cite{BAGP}. 
However, a smeared transition is expected again, since the densities in the 
steady state $\rho(t\to\infty)$ increase with $N$.
As the inset of Fig.~\ref{gwbasisl} shows $\rho(\lambda,t\to\infty)$ can be
fitted using
\begin{equation}\label{betafit}
\rho(\lambda,t\to\infty) = C (\lambda-\lambda_c)^{\beta} 
\end{equation}
with an order parameter exponent $\beta = 1.22(1)$ near $\lambda_c=1.64(1)$ 
for networks with size $N = 4\times 10^6$.

%%%%%%%%%%%%%%%%%%%%%%%%%%%%%%%%%%%%%%%%%%%%%%%%%%%%%%%%%%%%%%%%%%%%%%%%
\section{Discussion and Conclusions}
%%%%%%%%%%%%%%%%%%%%%%%%%%%%%%%%%%%%%%%%%%%%%%%%%%%%%%%%%%%%%%%%%%%%%%%%

Understanding effects of heterogeneities in nonequilibrium, dynamical systems
is a challenging open field. References \cite{GPCP,Juhasz:2011fk} 
concluded that finite topological dimension is a necessary condition 
for observing Griffiths Phases and activated scaling in case of the 
basic model of nonequilibrium system, the Contact Process. 
In case of CP on certain weighed networks numerical evidence was 
shown for generic power-laws, but in the thermodynamic limit this phase
seems to disappear and a smeared phase transition exists, due to the
infinite dimensional correlated rare-regions \cite{BAGP}.
In contrast with these strong disorder renormalization studies of the random 
transverse-field Ising model found Griffiths singularities \cite{IK11} even in 
the infinite dimensional Erd\H os-R\'enyi random graphs \cite{ER}.

Very recently Griffiths phases have been reported in a study of the 
Random Transverse Ising Model on complex networks with a scale-free 
degree distribution regularized by  and exponential cutoff 
$p(k) \propto k^{-\gamma}\exp[-k/\xi]$ \cite{Bianconi}. 
This model was devised to understand the relation between the onset of the 
superconducting state with the particular optimum heterogeneity in granular superconductors. On quenched networks a phase transition at zero temperature and a 
Griffith phase with decreasing size as the function of the cutoff $\xi$ was 
found. The scenario is similar to our case,because increasing the value 
of the cutoff is like going to the thermodynamic limit of the SF network.
Another fresh model study \cite{LSN12} argues again for the existence of 
GP in the SIS model defined on unclustered, deterministic SF networks, 
with $\gamma > 3$, below the percolation threshold of the network. 

In this work I provided spectral decomposition and QMF approximation 
to SIS models on different scale-free networks. This analysis has been 
supplemented with extensive numerical simulations showing dynamical 
effects of the topological disorder. 
Activity localization of the eigenvectors characterized by the IPR 
number predicts strong rare-region effects in the case of weighted 
WBAT-II trees. Numerical simulations exhibit a $\lambda$ parameter 
region, with continuously changing power-laws decays. Still, in the 
$N\to\infty$ limit the $\rho(t)$ curves saturate, as for smeared phase 
transitions, discussed in \cite{BAGP}. For the WBAT-I tree
the picture is less clear. There is a narrow GP like region at very 
late times ($t > 3\times10^5$ MCs), but the $\ln(\rho(\ln(t)))$ 
curves exhibit downward curvature and stretched exponential 
dynamics looks more reasonable. This scenario is strengthened by the 
QMF SD analysis, which results in a vanishing IPR, suggesting only 
weak rare-region effects here. One can accept this because the
neighbor infection in the SIS is more homogeneous than in the CP,
thus the disorder is weaker. This means that reducing the hub-hub 
weights is not enough in SIS to find strong rare region effects, 
but disassortativity can make the job. 

This study goes one step further than \cite{GDOM12} by analyzing spectral 
data of different graphs via finite size scaling. In particular numerical 
evidence is shown for the disappearance of $\lambda_c$ with the 
expected scaling law in the $N\to\infty$ limit.
Fluctuations, omitted by the QMF approximation can be relevant, 
indeed simulations of SIS on finite graphs show $\lambda_c>0$. 
However, finite size study suggest $\lambda_c=0$ in all cases, in 
accordance with smeared phase transition.
Still the QMF SD analysis seems to be a relatively fast, promising way 
to explore topological rare-region effects, GPs in networks, 
especially with infinite topological dimension. 
Extension of this method for using more powerful numerical techniques is 
under way.

\section*{Acknowledgments}

I thank R. Pastor-Satorras, R. Juh\'asz and I. Kov\'acs for useful 
discussions and acknowledge support from the Hungarian research
fund OTKA (Grant No. T77629), HPC-EUROPA2 (pr. 228398) and the
European Social Fund through project FuturICT.hu (grant no.:
TAMOP-4.2.2.C-11/1/KONV-2012-0013).


\begin{thebibliography}{50}%
\makeatletter
\providecommand \@ifxundefined [1]{%
 \@ifx{#1\undefined}
}%
\providecommand \@ifnum [1]{%
 \ifnum #1\expandafter \@firstoftwo
 \else \expandafter \@secondoftwo
 \fi
}%
\providecommand \@ifx [1]{%
 \ifx #1\expandafter \@firstoftwo
 \else \expandafter \@secondoftwo
 \fi
}%
\providecommand \natexlab [1]{#1}%
\providecommand \enquote  [1]{``#1''}%
\providecommand \bibnamefont  [1]{#1}%
\providecommand \bibfnamefont [1]{#1}%
\providecommand \citenamefont [1]{#1}%
\providecommand \href@noop [0]{\@secondoftwo}%
\providecommand \href [0]{\begingroup \@sanitize@url \@href}%
\providecommand \@href[1]{\@@startlink{#1}\@@href}%
\providecommand \@@href[1]{\endgroup#1\@@endlink}%
\providecommand \@sanitize@url [0]{\catcode `\\12\catcode `\$12\catcode
  `\&12\catcode `\#12\catcode `\^12\catcode `\_12\catcode `\%12\relax}%
\providecommand \@@startlink[1]{}%
\providecommand \@@endlink[0]{}%
\providecommand \url  [0]{\begingroup\@sanitize@url \@url }%
\providecommand \@url [1]{\endgroup\@href {#1}{\urlprefix }}%
\providecommand \urlprefix  [0]{URL }%
\providecommand \Eprint [0]{\href }%
\providecommand \doibase [0]{http://dx.doi.org/}%
\providecommand \selectlanguage [0]{\@gobble}%
\providecommand \bibinfo  [0]{\@secondoftwo}%
\providecommand \bibfield  [0]{\@secondoftwo}%
\providecommand \translation [1]{[#1]}%
\providecommand \BibitemOpen [0]{}%
\providecommand \bibitemStop [0]{}%
\providecommand \bibitemNoStop [0]{.\EOS\space}%
\providecommand \EOS [0]{\spacefactor3000\relax}%
\providecommand \BibitemShut  [1]{\csname bibitem#1\endcsname}%
\let\auto@bib@innerbib\@empty
%</preamble>
  \bibitem [{\citenamefont {Marro}\ and\ \citenamefont
  {Dickman}(1999)}]{marro1999npt}%
  \BibitemOpen
  \bibfield  {author} {\bibinfo {author} {\bibfnamefont {J.}~\bibnamefont
  {Marro}}\ and\ \bibinfo {author} {\bibfnamefont {R.}~\bibnamefont
  {Dickman}},\ }\href@noop {} {\emph {\bibinfo {title} {{Nonequilibrium Phase
  Transitions in Lattice Models}}}}\ (\bibinfo  {publisher} {Cambridge
  University Press},\ \bibinfo {address} {Cambridge},\ \bibinfo {year}
  {1999})\BibitemShut {NoStop}
  \bibitem [{\citenamefont {\'Odor}(2008)}]{odorbook}%
  \BibitemOpen
  \bibfield  {author} {\bibinfo {author} {\bibfnamefont {G.}~\bibnamefont
  {\'Odor}},\ }\href@noop {} {\emph {\bibinfo {title} {Universality in
  Nonequilibrium Lattice Systems}}}\ (\bibinfo  {publisher} {World
  Scientific},\ \bibinfo {address} {Singapore},\ \bibinfo {year}
  {2008})\BibitemShut {NoStop}%
  \bibitem [{\citenamefont {Henkel}\ \emph {et~al.}(2008)\citenamefont {Henkel},
  \citenamefont {Hinrichsen},\ and\ \citenamefont {L\"ubeck}}]{Henkel}%
  \BibitemOpen
  \bibfield  {author} {\bibinfo {author} {\bibfnamefont {M.}~\bibnamefont
  {Henkel}}, \bibinfo {author} {\bibfnamefont {H.}~\bibnamefont {Hinrichsen}},
  \ and\ \bibinfo {author} {\bibfnamefont {S.}~\bibnamefont {L\"ubeck}},\
  }\href@noop {} {\emph {\bibinfo {title} {Non-equilibrium phase transition:
  Absorbing Phase Transitions}}}\ (\bibinfo  {publisher} {Springer Verlag},\
  \bibinfo {address} {Netherlands},\ \bibinfo {year} {2008})\BibitemShut
  {NoStop}%
  \bibitem [{\citenamefont {Harris}(1974{\natexlab{a}})}]{harris74}%
  \BibitemOpen
  \bibfield  {author} {\bibinfo {author} {\bibfnamefont {T.~E.}\ \bibnamefont
  {Harris}},\ }\href@noop {} {\bibfield  {journal} {\bibinfo  {journal} {Ann.
  Prob.}\ }\textbf {\bibinfo {volume} {2}},\ \bibinfo {pages} {969} (\bibinfo
  {year} {1974}{\natexlab{a}})}\BibitemShut {NoStop}%
  \bibitem [{\citenamefont {Liggett}(1985)}]{liggett1985ips}%
  \BibitemOpen
  \bibfield  {author} {\bibinfo {author} {\bibfnamefont {T.~M.}\ \bibnamefont
  {Liggett}},\ }\href@noop {} {\emph {\bibinfo {title} {{Interacting Particle
  Systems}}}}\ (\bibinfo  {publisher} {Springer-Verlag},\ \bibinfo {address}
  {New York},\ \bibinfo {year} {1985})\BibitemShut {NoStop}%
  \bibitem [{\citenamefont {Albert}\ and\ \citenamefont
  {Barab\'asi}(2002)}]{barabasi02}%
  \BibitemOpen
  \bibfield  {author} {\bibinfo {author} {\bibfnamefont {R.}~\bibnamefont
  {Albert}}\ and\ \bibinfo {author} {\bibfnamefont {A.-L.}\ \bibnamefont
  {Barab\'asi}},\ }\href@noop {} {\bibfield  {journal} {\bibinfo  {journal}
  {Rev. Mod. Phys.}\ }\textbf {\bibinfo {volume} {74}},\ \bibinfo {pages} {47}
  (\bibinfo {year} {2002})}\BibitemShut {NoStop}%
  \bibitem [{\citenamefont {Dorogovtsev}\ and\ \citenamefont
  {Mendes}(2003)}]{mendesbook}%
  \BibitemOpen
  \bibfield  {author} {\bibinfo {author} {\bibfnamefont {S.~N.}\ \bibnamefont
  {Dorogovtsev}}\ and\ \bibinfo {author} {\bibfnamefont {J.~F.~F.}\
  \bibnamefont {Mendes}},\ }\href@noop {} {\emph {\bibinfo {title} {Evolution
  of networks: From biological nets to the {I}nternet and {WWW}}}}\ (\bibinfo
  {publisher} {Oxford University Press},\ \bibinfo {address} {Oxford},\
  \bibinfo {year} {2003})\BibitemShut {NoStop}%
  \bibitem [{\citenamefont {Dorogovtsev}\ \emph {et~al.}(2008)\citenamefont
  {Dorogovtsev}, \citenamefont {Goltsev},\ and\ \citenamefont
  {Mendes}}]{dorogovtsev07:_critic_phenom}%
  \BibitemOpen
  \bibfield  {author} {\bibinfo {author} {\bibfnamefont {S.~N.}\ \bibnamefont
  {Dorogovtsev}}, \bibinfo {author} {\bibfnamefont {A.~V.}\ \bibnamefont
  {Goltsev}}, \ and\ \bibinfo {author} {\bibfnamefont {J.~F.~F.}\ \bibnamefont
  {Mendes}},\ }\href@noop {} {\bibfield  {journal} {\bibinfo  {journal} {Rev.
  Mod. Phys.}\ }\textbf {\bibinfo {volume} {80}},\ \bibinfo {pages} {1275}
  (\bibinfo {year} {2008})}\BibitemShut {NoStop}%
  \bibitem [{\citenamefont {Barrat}\ \emph {et~al.}(2008)\citenamefont {Barrat},
  \citenamefont {Barth\'{e}lemy},\ and\ \citenamefont
  {Vespignani}}]{barratbook}%
  \BibitemOpen
  \bibfield  {author} {\bibinfo {author} {\bibfnamefont {A.}~\bibnamefont
  {Barrat}}, \bibinfo {author} {\bibfnamefont {M.}~\bibnamefont
  {Barth\'{e}lemy}}, \ and\ \bibinfo {author} {\bibfnamefont {A.}~\bibnamefont
  {Vespignani}},\ }\href@noop {} {\emph {\bibinfo {title} {Dynamical Processes
  on Complex Networks}}}\ (\bibinfo  {publisher} {Cambridge University Press},\
  \bibinfo {address} {Cambridge},\ \bibinfo {year} {2008})\BibitemShut
  {NoStop}%
\bibitem [{\citenamefont {Barab{\'a}si}\ and\ \citenamefont
  {Albert}(1999)}]{Barabasi:1999}%
  \BibitemOpen
  \bibfield  {author} {\bibinfo {author} {\bibfnamefont {A.-L.}\ \bibnamefont
  {Barab{\'a}si}}\ and\ \bibinfo {author} {\bibfnamefont {R.}~\bibnamefont
  {Albert}},\ }\href@noop {} {\bibfield  {journal} {\bibinfo  {journal}
  {Science}\ }\textbf {\bibinfo {volume} {286}},\ \bibinfo {pages} {509}
  (\bibinfo {year} {1999})}\BibitemShut {NoStop}%
  (Oxford University Press, Oxford, 1992).
\bibitem [{\citenamefont {Castellano}\ and\ \citenamefont
  {Pastor-Satorras}(2006)}]{Castellano:2006}%
  \BibitemOpen
  \bibfield  {author} {\bibinfo {author} {\bibfnamefont {C.}~\bibnamefont
  {Castellano}}\ and\ \bibinfo {author} {\bibfnamefont {R.}~\bibnamefont
  {Pastor-Satorras}},\ }\href@noop {} {\bibfield  {journal} {\bibinfo
  {journal} {Phys. Rev. Lett.}\ }\textbf {\bibinfo {volume} {96}},\ \bibinfo
  {pages} {038701} (\bibinfo {year} {2006})}\BibitemShut {NoStop}%
\bibitem [{\citenamefont {Castellano}\ and\ \citenamefont
  {Pastor-Satorras}(2008)}]{Castellano:2008}%
  \BibitemOpen
  \bibfield  {author} {\bibinfo {author} {\bibfnamefont {C.}~\bibnamefont
  {Castellano}}\ and\ \bibinfo {author} {\bibfnamefont {R.}~\bibnamefont
  {Pastor-Satorras}},\ }\href@noop {} {\bibfield  {journal} {\bibinfo
  {journal} {Phys. Rev. Lett.}\ }\textbf {\bibinfo {volume} {100}},\ \bibinfo
  {pages} {148701} (\bibinfo {year} {2008})}\BibitemShut {NoStop}%
\bibitem [{\citenamefont {Ferreira}\ \emph
  {et~al.}(2011{\natexlab{a}})\citenamefont {Ferreira}, \citenamefont
  {Ferreira},\ and\ \citenamefont {Pastor-Satorras}}]{PhysRevE.83.066113}%
  \BibitemOpen
  \bibfield  {author} {\bibinfo {author} {\bibfnamefont {S.~C.}\ \bibnamefont
  {Ferreira}}, \bibinfo {author} {\bibfnamefont {R.~S.}\ \bibnamefont
  {Ferreira}}, \ and\ \bibinfo {author} {\bibfnamefont {R.}~\bibnamefont
  {Pastor-Satorras}},\ }\href@noop {} {\bibfield  {journal} {\bibinfo
  {journal} {Phys. Rev. E}\ }\textbf {\bibinfo {volume} {83}},\ \bibinfo
  {pages} {066113} (\bibinfo {year} {2011}{\natexlab{a}})}\BibitemShut
  {NoStop}%
\bibitem [{\citenamefont {Ferreira}\ \emph
  {et~al.}(2011{\natexlab{b}})\citenamefont {Ferreira}, \citenamefont
  {Ferreira}, \citenamefont {Castellano},\ and\ \citenamefont
  {Pastor-Satorras}}]{FFCR11}%
  \BibitemOpen
  \bibfield  {author} {\bibinfo {author} {\bibfnamefont {S.~C.}\ \bibnamefont
  {Ferreira}}, \bibinfo {author} {\bibfnamefont {R.~S.}\ \bibnamefont
  {Ferreira}}, \bibinfo {author} {\bibfnamefont {C.}~\bibnamefont
  {Castellano}}, \ and\ \bibinfo {author} {\bibfnamefont {R.}~\bibnamefont
  {Pastor-Satorras}},\ }\href@noop {} {\bibfield  {journal} {\bibinfo
  {journal} {Phys. Rev. E}\ }\textbf {\bibinfo {volume} {84}},\ \bibinfo
  {pages} {066102} (\bibinfo {year} {2011}{\natexlab{b}})}\BibitemShut
  {NoStop}%
\bibitem [{\citenamefont {{Bogu\~{n}\'{a}}}\ \emph {et~al.}(2009)\citenamefont
  {{Bogu\~{n}\'{a}}}, \citenamefont {Castellano},\ and\ \citenamefont
  {Pastor-Satorras}}]{boguna09:_langev}%
  \BibitemOpen
  \bibfield  {author} {\bibinfo {author} {\bibfnamefont {M.}~\bibnamefont
  {{Bogu\~{n}\'{a}}}}, \bibinfo {author} {\bibfnamefont {C.}~\bibnamefont
  {Castellano}}, \ and\ \bibinfo {author} {\bibfnamefont {R.}~\bibnamefont
  {Pastor-Satorras}},\ }\href@noop {} {\bibfield  {journal} {\bibinfo
  {journal} {Phys. Rev. E}\ }\textbf {\bibinfo {volume} {79}},\ \bibinfo
  {pages} {036110} (\bibinfo {year} {2009})}\BibitemShut {NoStop}%
\bibitem [{\citenamefont {Mu\~noz}\ \emph {et~al.}(2010)\citenamefont
  {Mu\~noz}, \citenamefont {Juh\'asz}, \citenamefont {Castellano},\ and\
  \citenamefont {\'Odor}}]{GPCP}
  \BibitemOpen
  \bibfield  {author} {\bibinfo {author} {\bibfnamefont {M.~A.}\ \bibnamefont
  {Mu\~noz}}, \bibinfo {author} {\bibfnamefont {R.}~\bibnamefont {Juh\'asz}},
  \bibinfo {author} {\bibfnamefont {C.}~\bibnamefont {Castellano}}, \ and\
  \bibinfo {author} {\bibfnamefont {G.}~\bibnamefont {\'Odor}},\ }\href@noop {}
  {\bibfield  {journal} {\bibinfo  {journal} {Phys. Rev. Lett.}\ }\textbf
  {\bibinfo {volume} {105}},\ \bibinfo {pages} {128701} (\bibinfo {year}
  {2010})}\BibitemShut {NoStop}%
\bibitem [{\citenamefont {\'Odor}\ \emph {et~al.}(2011)\citenamefont {\'Odor},
  \citenamefont {Juhasz}, \citenamefont {Castellano},\ and\ \citenamefont
  {Munoz}}]{odor:172}%
  \BibitemOpen
  \bibfield  {author} {\bibinfo {author} {\bibfnamefont {G.}~\bibnamefont
  {\'Odor}}, \bibinfo {author} {\bibfnamefont {R.}~\bibnamefont {Juhasz}},
  \bibinfo {author} {\bibfnamefont {C.}~\bibnamefont {Castellano}}, \ and\
  \bibinfo {author} {\bibfnamefont {M.~A.}\ \bibnamefont {Munoz}},\ }in\
  \href@noop {} {\emph {\bibinfo {booktitle} {Nonequilibrium Statistical
  Physics Today}}},\ Vol.\ \bibinfo {volume} {1332},\ \bibinfo {editor} {edited
  by\ \bibinfo {editor} {\bibfnamefont {P.~L.}\ \bibnamefont {Garrido}},
  \bibinfo {editor} {\bibfnamefont {J.}~\bibnamefont {Marro}}, \ and\ \bibinfo
  {editor} {\bibfnamefont {F.}~\bibnamefont {de~los Santos}}}\ (\bibinfo
  {publisher} {AIP},\ \bibinfo {year} {2011})\ pp.\ \bibinfo {pages}
  {172--178}\BibitemShut {NoStop}%
\bibitem [{\citenamefont {{Juh{\'a}sz}}\ \emph {et~al.}(2011)\citenamefont
  {{Juh{\'a}sz}}, \citenamefont {{{\'O}dor}}, \citenamefont {{Castellano}},\
  and\ \citenamefont {{Mu{\~n}oz}}}]{Juhasz:2011fk}%
  \BibitemOpen
  \bibfield  {author} {\bibinfo {author} {\bibfnamefont {R.}~\bibnamefont
  {{Juh{\'a}sz}}}, \bibinfo {author} {\bibfnamefont {G.}~\bibnamefont
  {{{\'O}dor}}}, \bibinfo {author} {\bibfnamefont {C.}~\bibnamefont
  {{Castellano}}}, \ and\ \bibinfo {author} {\bibfnamefont {M.~A.}\
  \bibnamefont {{Mu{\~n}oz}}},\ \href@noop {}
  {\bibfield  {journal} {\bibinfo  {journal} {Phys. Rev. E.}\ }\textbf
  {\bibinfo {volume} {85}},\ \bibinfo {pages} {066125} (\bibinfo {year}
  {2012})}\BibitemShut {NoStop}}
\bibitem{Johnson} S. Johnson, J. J. Torres, and J. Marro, PLoS ONE 
{\bf 8(1)}: e50276 (2013)
\bibitem{Chialvo} D. R. Chialvo, {\it Criticality in Neural Systems},
 Niebur E, Plenz D, Schuster HG. (eds.) {\it John Wiley \& Sons} (2013), arXiv:1210.3632 
\bibitem{Castello} X. Castell\'o, R. Toivonen, V. M. Egu\'\i luz, J. Saram\"aki, 
K. Kaski and M. San Miguel, EPL {\bf 79} (2007) 66006.
\bibitem{Amir} A. Amir, Y. Oreg, and Y. Imry, Phys. Rev. Lett. {\bf 105}, 070601 (2010).
\bibitem{KK11} M. Karsai, et al. Phys. Rev. E. {\bf 83}, 025102(R) (2011).
\bibitem [{\citenamefont {Griffiths}(1969)}]{Griffiths}%
  \BibitemOpen
  \bibfield  {author} {\bibinfo {author} {\bibfnamefont {R.~B.}\ \bibnamefont
  {Griffiths}},\ }\href@noop {} {\bibfield  {journal} {\bibinfo  {journal}
  {Phys. Rev. Lett.}\ }\textbf {\bibinfo {volume} {23}},\ \bibinfo {pages} {17}
  (\bibinfo {year} {1969})}\BibitemShut {NoStop}%
\bibitem [{\citenamefont {Vojta}(2006)}]{Vojta}%
  \BibitemOpen
  \bibfield  {author} {\bibinfo {author} {\bibfnamefont {T.}~\bibnamefont
  {Vojta}},\ }\href@noop {} {\bibfield  {journal} {\bibinfo  {journal} {Journal
  of Physics A: Mathematical and General}\ }\textbf {\bibinfo {volume} {39}},\
  \bibinfo {pages} {R143} (\bibinfo {year} {2006})}\BibitemShut {NoStop}%
\bibitem [{\citenamefont {Bray}(1987)}]{PhysRevLett.59.586}%
  \BibitemOpen
  \bibfield  {author} {\bibinfo {author} {\bibfnamefont {A.~J.}\ \bibnamefont
  {Bray}},\ }\href@noop {} {\bibfield  {journal} {\bibinfo  {journal} {Phys.
  Rev. Lett.}\ }\textbf {\bibinfo {volume} {59}},\ \bibinfo {pages} {586}
  (\bibinfo {year} {1987})}\BibitemShut {NoStop}%
\bibitem [{\citenamefont {Dhar}\ \emph {et~al.}(1988)\citenamefont {Dhar},
  \citenamefont {Randeria},\ and\ \citenamefont {Sethna}}]{sethna88}%
  \BibitemOpen
  \bibfield  {author} {\bibinfo {author} {\bibfnamefont {D.}~\bibnamefont
  {Dhar}}, \bibinfo {author} {\bibfnamefont {M.}~\bibnamefont {Randeria}}, \
  and\ \bibinfo {author} {\bibfnamefont {J.~P.}\ \bibnamefont {Sethna}},\
  }\href@noop {} {\bibfield  {journal} {\bibinfo  {journal} {Europhys. Lett.}\
  }\textbf {\bibinfo {volume} {5}},\ \bibinfo {pages} {485} (\bibinfo {year}
  {1988})}\BibitemShut {NoStop}%
\bibitem [{\citenamefont {Rieger}\ and\ \citenamefont
  {Young}(1996)}]{PhysRevB.54.3328}%
  \BibitemOpen
  \bibfield  {author} {\bibinfo {author} {\bibfnamefont {H.}~\bibnamefont
  {Rieger}}\ and\ \bibinfo {author} {\bibfnamefont {A.~P.}\ \bibnamefont
  {Young}},\ }\href@noop {} {\bibfield  {journal} {\bibinfo  {journal} {Phys.
  Rev. B}\ }\textbf {\bibinfo {volume} {54}},\ \bibinfo {pages} {3328}
  (\bibinfo {year} {1996})}\BibitemShut {NoStop}%
\bibitem [{\citenamefont {Fisher}(1992)}]{PhysRevLett.69.534}%
  \BibitemOpen
  \bibfield  {author} {\bibinfo {author} {\bibfnamefont {D.~S.}\ \bibnamefont
  {Fisher}},\ }\href@noop {} {\bibfield  {journal} {\bibinfo  {journal} {Phys.
  Rev. Lett.}\ }\textbf {\bibinfo {volume} {69}},\ \bibinfo {pages} {534}
  (\bibinfo {year} {1992})}\BibitemShut {NoStop}%
\bibitem{BAGP} G.\'Odor and R. Pastor-Satorras, Phys. Rev. E {\bf 86}, (2012) 026117.
\bibitem{SIS} R. M. Anderson and R. M. May, {\it Infectious diseases in humans},
  (Oxford University Press, Oxford, 1992).
\bibitem [{\citenamefont {Catanzaro}\ \emph
  {et~al.}(2005{\natexlab{a}})\citenamefont {Catanzaro}, \citenamefont
  {{Bogu\~{n}\'{a}}},\ and\ \citenamefont {Pastor-Satorras}}]{ucmmodel}%
  \BibitemOpen
  \bibfield  {author} {\bibinfo {author} {\bibfnamefont {M.}~\bibnamefont
  {Catanzaro}}, \bibinfo {author} {\bibfnamefont {M.}~\bibnamefont
  {{Bogu\~{n}\'{a}}}}, \ and\ \bibinfo {author} {\bibfnamefont
  {R.}~\bibnamefont {Pastor-Satorras}},\ }\href@noop {} {\bibfield  {journal}
  {\bibinfo  {journal} {Phys. Rev. E}\ }\textbf {\bibinfo {volume} {71}},\
  \bibinfo {pages} {027103} (\bibinfo {year} {2005}{\natexlab{a}})}\BibitemShut
  {NoStop}%
\bibitem{GDOM12} A. V. Goltsev, S. N. Dorogovtsev, J. G. Oliveira,
        and J. F. F.  Mendes, Phys. Rev. Lett. {\bf 109}, 128702 (2012).
\bibitem{Mie} P. Van Mieghem, Eur. Phys. Lett. {\bf 97}, 48004 (2012).
\bibitem{Chung} F. Chung, L. Lu and V. Vu, Proc. Natl. Acad. Sci. USA {\bf 100} 6313 (2003)
\bibitem{Heap} H. Heaps, {\it Information retrieval: Computational and theoretical aspects} (Academic Press, Inc. Orlando, FL, USA, 1978).
\bibitem{FCP12} S. C. Ferreira, C. Castellano, R. Pastor-Satorras, 
  Phys. Rev. E {\bf 86}, 041125 (2012)
\bibitem{Karsai} M. Karsai, R. Juh\'asz, and F. Igl\'oi, 
    Phys. Rev. E {\bf 73}, 036116 (2006).
\bibitem{Bianconi} G. Bianconi, J. Stat. Mech. (2012) P07021.
\bibitem{LSN12} H. K. Lee, P.-S. Shim and J. D. Noh, arXiv:1211.2519.
\bibitem{IK11} I. A. Kov\'acs, F. Igl\'oi, J. Phys.: Condens. Matter {\bf 23} (2011) 404204.
\bibitem{ER} P. Erd\H os, A. R\'enyi, Publicationes Mathematicae {\bf 6}, 290 (1959).
 

\end{thebibliography}
\end{document}